\documentclass[10pt]{article}

\usepackage[top=24mm, bottom=24mm, left=30mm, right=30mm]{geometry}
\usepackage{url, graphicx, color, varioref, amsfonts, longtable, float}

\newcommand{\qed}{\nobreak \ifvmode \relax \else
      \ifdim\lastskip<1.5em \hskip-\lastskip
      \hskip1.5em plus0em minus0.5em \fi \nobreak
      \vrule height0.75em width0.5em depth0.25em\fi}

\title{Database Manipulation on Quantum Computers}

\author{Ahmed Younes\footnote {ayounes2@yahoo.com}\\
Department of Math. \& Comp. Science\\
Faculty of Science\\
Alexandria University\\
Alexandria, Egypt}

\begin{document}
\maketitle
\begin{abstract}
 Manipulating a database system on a quantum computer is an essential aim to benefit from the promising speed-up of quantum computers 
over classical computers in areas that take a vast amount of storage and processing time such as in databases. 
In this paper, the basic operations for manipulating the data in a quantum database will be defined, 
e.g. INSERT, UPDATE, DELETE, SELECT, backing up and restoring a database file. 
This gives the ability to perform the data processing, that usually takes a long processing time on a 
classical database system, in a simultaneous way on a quantum computer. 
Defining a quantum version of more advanced concepts used in database systems, 
e.g. the referential integrity and the relational algebra, 
is a normal extension to this work.
\end{abstract}

\section{Introduction}       
Quantum computers promise to do computation more powerfully \cite{simon94} 
than classical computers due to the ability of a quantum computer to be 
in some states that have no equivalence in a classical 
computer such as a superposition of values and/or an entanglement 
between some particles of a quantum system \cite{Feynman82}. 
A superposition is 
the ability to have more than one value stored simultaneously over the same physical space while an 
entanglement is the existence of a hidden correlation between the particles of a quantum system 
\cite{Bell66} so that applying an operation on an entangled particle will apply that operation on 
all the particles entangled with that particle \cite{Bennett93}. A quantum computer 
exploits a superposition to perform parallel computation on many values simultaneously at the bit level 
while a classical computer can perform simultaneous operations at the CPU level \cite{Rieffel00}.

To extract information from a quantum computer, a system measurement must be used \cite{Rieffel00}.
If that quantum computer exists in a superposition, the measurement will break 
the superposition to one of the superposed values in a random manner. Otherwise, 
a quantum computer behaves classically, i.e. if no superposition exists. 
Many useful methods are known to increase the probability of a required value to be found 
with a probability close to certainty when the measurement is applied \cite{Brassard98,Mosca98,Brassard00,boyer96,younes-pi1825}. 

Many quantum algorithms exploit a superposition and/or an entanglement to perform computation 
faster than it can be done on classical computers \cite{shor94,grover96,Younes03d}, 
where all the possible inputs of a problem are examined simultaneously. 
A superposition can be understood as a list of values superposed together on the same memory location. 
A database file is a two dimensional data structure (a table) where every column represents a field over 
certain data type and every row represents a record (a collection of related fields) \cite{DBMS}. 
A database file is simply a list of unique records. Combining the fields in each record in 
some fixed binary representation, a list of records can be manipulated as a list of values that can 
exist in a superposition on a quantum computer.

Structured Query Language (SQL) is a tool widely used in manipulating the classical databases \cite{DBMS}. 
Basic operations in SQL include inserting a new record to a database file (INSERT), updating an existing record (UPDATE), 
deleting an exiting record (DELETE), selecting (SELECT) and performing 
an arbitrary operation on some records, backing up a portion of a database (BACKUP), and restoring the backup (RESTORE). 
In this paper, elementary operations for a Quantum Query Language (QQL) required to manipulate 
a database file exists in a superposition will be defined. 

The paper is organized as follows: 
Section 2 briefly reviews the basic concepts in quantum computation. Section 3 defines the basic quantum transformations 
required to construct the QQL. Section 4 defines the basic operators of the QQL. 
Section 5 will conclude the work showing some future directions to the way of 
constructing a complete Quantum Database Management System (QDBMS).  

\section{Quantum Computers}

\subsection{Quantum Bits}

The quantum bit ({\it qubit} \cite{sch95}) is the quantum analogue of the classical bit. The basic 
difference between the qubit and the classical bit is that the qubit can exist in a 
linear superposition of the two states $\left| 0 \right\rangle$ and  $\left| 1 \right\rangle$ at the same time 
({\it Quantum Parallelism}),
\begin{equation}
a\left| 0 \right\rangle  +b \left| 1 \right\rangle,
\end{equation}
where $a$ and $b$ are complex numbers called the amplitudes of system and satisfy 
the condition $\left| a \right|^2  + \left| b \right|^2  = 1$. The states $\left| 0 \right\rangle $ 
and $\left| 1 \right\rangle $ can be taken as the classical bit values 0 and 1 respectively. 
$\left| {\,\,\,\,} \right\rangle$ is called the {\it Dirac notation} \cite{dirac47a} and is considered as the 
standard notation for describing quantum states. In quantum circuits shown in this paper, 
a qubit will be represented as a horizontal line and the time flow of the circuit will be from left to right. 

\section{Multiple Qubits}

Consider the case where we have a quantum system (quantum register) with more than one qubit. 
In conventional computers, a two-bit register will be able to carry only one value out of the four 
possible values $\{00, 01, 10, 11\}$ at a time. The corresponding states in a two-qubit quantum register will be 
$\{\left| {00} \right\rangle$, $\left| {01} \right\rangle$, $\left| {10} \right\rangle$, $\left| {11} \right\rangle \}$, 
so its state in a superposition can be represented as,
\begin{equation}
\left| \psi  \right\rangle  = a_0 \left| {00} \right\rangle  + a_1 \left| {01} \right\rangle  + a_2 \left| {10} \right\rangle  + a_3 \left| {11} \right\rangle,
\label{ch2eqn2.11}
\end{equation}
\noindent
where $a_i$ are complex numbers satisfy the condition $\sum\nolimits_i {{|a_i|}}^2=1$. Any measurement 
applied on the qubits will lead to one of the four 
possible states $\left|i\right\rangle$ with probability $\left| {a_i } \right|^2$, 
where $i$ is the integer representation of that state. 

Before we go further, it is important to review some useful mathematical concepts \cite{nc00a,Rieffel00}: 
The state of $n$-qubit quantum system can be represented as a vector of length $2^n$ over {\it Hilbert space}. 
States can be represented via either the vector/matrix notation, or Dirac Notation (Ket/Bra notation)\cite{dirac47a}.
Dirac Notation is more useful for describing the quantum states and the evolution of the state of the system, 
it can be understood as follows:

\begin{itemize}

\item{\it Ket} 
$\left| \psi  \right\rangle$: denotes a column vector that represents a quantum state.

\item{\it Bra} 
$\left\langle \psi  \right|$: denotes a row vector that represents the dual of the ket, i.e. the complex conjugate 
transpose of $\left| \psi  \right\rangle$. 

\item{\it The inner product of two vectors} is written as 
$\left\langle \psi  \right|\left| \xi  \right\rangle $ or shortly $\left\langle {\psi }
 \mathrel{\left | {\vphantom {\psi  \xi }}
 \right. \kern-\nulldelimiterspace}
 {\xi } \right\rangle $. Notice that, since $\left| 0 \right\rangle$
 is a unit vector, we have $\left\langle {0}
 \mathrel{\left | {\vphantom {0 0}}
 \right. \kern-\nulldelimiterspace}
 {0} \right\rangle  = 1$ and since $\left| 0 \right\rangle$
 and $\left| 1 \right\rangle$ are orthogonal, we have $\left\langle {0}
 \mathrel{\left | {\vphantom {0 1}}
 \right. \kern-\nulldelimiterspace}
 {1} \right\rangle  = 0$. 

\item{\it The outer product of two vectors} 
is written as $\left| \psi  \right\rangle \left\langle \xi  \right|$. 
A matrix (operator) can be represented in the outer product form, where it is sometimes called the {\it diagonal 
representation} of that operator. For example, the Identity gate can be represented as follows,

\begin{equation}
I = \left| 0 \right\rangle \left\langle 0 \right| + \left| 1 \right\rangle \left\langle 1 \right| = \left[ {\begin{array}{*{20}c}
   1 & 0  \\
   0 & 1  \\
\end{array}} \right].
\end{equation}

\item{\it The tensor product of two vectors} $\left| \psi  \right\rangle$ and $\left| \xi  \right\rangle$ 
is written as $\left|\psi\right\rangle\otimes\left|\xi\right\rangle$ and is
used to combine smaller quantum systems in a single larger quantum system. For example, let $\left|\psi\right\rangle$ 
and $\left|\xi\right\rangle$ be vectors from a two-dimensional complex vector space 
spanned by the basis $\{\left| 0 \right\rangle,\left| 1 \right\rangle\}$. 
The tensor product of $\left|\psi\right\rangle$ and $\left|\xi\right\rangle$ will have the basis,
 
\begin{equation}
\left( {\left|0\right\rangle  \otimes \left|0\right\rangle ,\left|0\right\rangle  \otimes \left|1\right\rangle ,\left|1\right\rangle
  \otimes \left|0\right\rangle ,\left|1\right\rangle  \otimes \left|1\right\rangle } \right),
\end{equation}

\noindent
where the order of the basis is arbitrary as long as it is fixed, which can be re-written shortly as,
\begin{equation}
\left( {\left| {00} \right\rangle ,\left| {01} \right\rangle ,\left| {10} \right\rangle ,\left| {11} \right\rangle } \right).
\label{ch2eqn2.12}
\end{equation}
 
Similarly, basis for a three-qubit system will be,

\begin{equation}
\left( {\left| {000} \right\rangle ,\left| {001} \right\rangle ,\left| {010} \right\rangle ,\left| {011} \right\rangle ,\left| {100} \right\rangle ,\left| {101} \right\rangle ,\left| {110} \right\rangle ,\left| {111} \right\rangle } \right).
\label{ch2eqn2.13}
\end{equation}

Now, we can view the state of a single-qubit as a vector in the two-dimensional complex vector space spanned by 
the orthonormal basis $\{{\left| 0 \right\rangle}$ , ${\left| 1 \right\rangle}\}$ as follows, 

\begin{equation}
a \left| 0 \right\rangle  + b \left| 1 \right\rangle  = \left[ {\begin{array}{*{20}c}
   a   \\
   b   \\
\end{array}} \right],
\end{equation}

\noindent
where,

\begin{equation}
\left| 0 \right\rangle  = \left[ {\begin{array}{*{20}c}
   1  \\
   0  \\
\end{array}} \right],\left| 1 \right\rangle  = \left[ {\begin{array}{*{20}c}
   0  \\
   1  \\
\end{array}} \right].
\end{equation}

Similarly, the state of a two-qubit quantum register is a vector in the four-dimensional complex vector space spanned by 
the orthonormal basis $\{{\left| 00 \right\rangle}$, ${\left| 01 \right\rangle}$, ${\left| 10 \right\rangle}$, ${\left| 11 \right\rangle}\}$ as follows,

\begin{equation}
a_0 \left| {00} \right\rangle  + a_1 \left| {01} \right\rangle  + a_2 \left| {10} \right\rangle  + a_3 \left| {11} \right\rangle  = \left[ {\begin{array}{*{20}c}
   {a_0 }  \\
   {a_1 }  \\
   {a_2 }  \\
   {a_3 }  \\
\end{array}} \right],
\end{equation}

\noindent
where,

\begin{equation}
\left| {00} \right\rangle  = \left[ {\begin{array}{*{20}c}
   1  \\
   0  \\
   0  \\
   0  \\
\end{array}} \right],\,\,\left| {01} \right\rangle  = \left[ {\begin{array}{*{20}c}
   0  \\
   1  \\
   0  \\
   0  \\
\end{array}} \right],\,\,\left| {10} \right\rangle  = \left[ {\begin{array}{*{20}c}
   0  \\
   0  \\
   1  \\
   0  \\
\end{array}} \right],\,\,\left| {11} \right\rangle  = \left[ {\begin{array}{*{20}c}
   0  \\
   0  \\
   0  \\
   1  \\
\end{array}} \right].
\end{equation}

For a quantum system of $n$ qubits, the resulting state space 
is of dimension $2^n$. If the qubits of this quantum system are all initialised to the same state, for example, 
state $\left|0\right\rangle$, it can be written shortly as 
$\left| {00\ldots0} \right\rangle  = \left| 0 \right\rangle ^{ \otimes n}$. 
This exponential growth of the state space with the linear increase in 
the number of qubits is one of the reasons for the possibility of an exponential 
increase in the speed of computation on quantum computers over classical computers \cite{Rieffel00}.

\item{\it The tensor product of two operators} $U$ and $V$ 
is written as $U\otimes V$ and is
used to combine smaller quantum operators in a single larger operator. For example, let $U$ and $V$ to be 
single-qubit operators ($2 \times 2$ matrices) defined as follows,

\begin{equation}
U = \left[ {\begin{array}{*{20}c}
   {u_{00} } & {u_{01} }  \\
   {u_{10} } & {u_{11} }  \\
\end{array}} \right],\,\,\,\,V = \left[ {\begin{array}{*{20}c}
   {v_{00} } & {v_{01} }  \\
   {v_{10} } & {v_{11} }  \\
\end{array}} \right].
\end{equation}

Consider a two-qubit system $\left|\psi\right\rangle\otimes\left|\xi\right\rangle$. Applying $U$ on $\left|\psi\right\rangle$ 
and $V$ on $\left|\xi\right\rangle$ in parallel can be written as follows,

\begin{equation}
U \otimes V\left( {\left| \psi  \right\rangle  \otimes \left| \xi  \right\rangle } \right) = U\left| \psi  \right\rangle  \otimes V\left| \xi  \right\rangle.
\end{equation}
where $U \otimes V$ can be combined in a single matrix of size $4 \times 4$ as follows,

\begin{equation}
\begin{array}{l}
 U \otimes V = \left[ {\begin{array}{*{20}c}
   {u_{00} } & {u_{01} }  \\
   {u_{10} } & {u_{11} }  \\
\end{array}} \right] \otimes \left[ {\begin{array}{*{20}c}
   {v_{00} } & {v_{01} }  \\
   {v_{10} } & {v_{11} }  \\
\end{array}} \right] \\
   \\ 
 \,\,\,\,\,\,\,\,\,\,\,\,\,\,\, = \left[ {\begin{array}{*{20}c}
   {u_{00} \left[ {\begin{array}{*{20}c}
   {v_{00} } & {v_{01} }  \\
   {v_{10} } & {v_{11} }  \\
\end{array}} \right]} & {u_{01} \left[ {\begin{array}{*{20}c}
   {v_{00} } & {v_{01} }  \\
   {v_{10} } & {v_{11} }  \\
\end{array}} \right]}  \\
   {u_{10} \left[ {\begin{array}{*{20}c}
   {v_{00} } & {v_{01} }  \\
   {v_{10} } & {v_{11} }  \\
\end{array}} \right]} & {u_{11} \left[ {\begin{array}{*{20}c}
   {v_{00} } & {v_{01} }  \\
   {v_{10} } & {v_{11} }  \\
\end{array}} \right]}  \\
\end{array}} \right] \\ 
\\
 \,\,\,\,\,\,\,\,\,\,\,\,\,\,\, = \left[ {\begin{array}{*{20}c}
   {\begin{array}{*{20}c}
   {u_{00} v_{00} } & {u_{00} v_{01} }  \\
   {u_{00} v_{10} } & {u_{00} v_{11} }  \\
\end{array}} & {\begin{array}{*{20}c}
   {u_{01} v_{00} } & {u_{01} v_{01} }  \\
   {u_{01} v_{10} } & {u_{01} v_{11} }  \\
\end{array}}  \\
   {\begin{array}{*{20}c}
   {u_{10} v_{00} } & {u_{10} v_{01} }  \\
   {u_{10} v_{10} } & {u_{10} v_{11} }  \\
\end{array}} & {\begin{array}{*{20}c}
   {u_{11} v_{00} } & {u_{11} v_{01} }  \\
   {u_{11} v_{10} } & {u_{11} v_{11} }  \\
\end{array}}  \\
\end{array}} \right]. \\ 
 \end{array}
\end{equation}

If $U$ is, for example, a $2\times2$ matrix and is tensored by itself $n$ times, so it can be written shortly as
$U \otimes U \otimes ... \otimes U = U^{ \otimes n}$, where the resulting matrix will be of size $2^n\times2^n$. 
More details on tensor products and their properties can be found in \cite{Hford,Rieffel00,tens}.

\end{itemize}

\subsection{Quantum Gates}

In general, quantum computation process can be understood as applying a series of quantum gates followed 
by applying a measurement to obtain the result \cite{nc00a}. Quantum gates used during the computation must follow the fundamental 
laws of quantum physics \cite{dirac47a}. To satisfy this condition, using any matrix $U$ as a quantum gate, 
it must be unitary, i.e. the inverse of that matrix must be equal to its complex conjugate transpose: 
$U^{ - 1} = U^\dag$ and $UU^\dag=I$, where $U^{-1}$ denotes the inverse of $U$, $U^\dag$ denotes the complex 
conjugate transpose of $U$ and $I$ is the identity matrix. Any gate applied on a quantum register of size $n$ 
can be understood by its action on the basis vectors and can be represented as a unitary matrix of size 
$2^n\times 2^n$.

For example, the $NOT$ gate is a 
single input/output gate that inverts the state $\left| 0 \right\rangle $ to $\left| 1 \right\rangle$ and 
visa versa. Its $2\times2$ unitary matrix: $NOT = \left[ {{\begin{array}{*{20}c}
 0 \hfill & 1 \hfill \\
 1 \hfill & 0 \hfill \\
\end{array} }} \right]$. Another important example is the Hadamard gate ($H$ gate) 
which produces a completely random output with equal 
probabilities to be $\left| 0 \right\rangle $ or $\left| 1 \right\rangle$ 
at any measurement. Its $2\times2$ unitary matrix: $H = \frac{1}{\sqrt 2 }\left[ {{
\begin{array}{*{20}c}
 1 \hfill & \,\,\,\,1 \hfill \\
 1 \hfill & { - 1} \hfill \\
\end{array}
}} \right]$. Hadamard gate has a special importance in setting up a superposition of a quantum register. 
Consider a three qubits quantum register $\left| 000 \right\rangle$, applying Hadamard gate on each of them 
in parallel will set up a superposition of the $2^3$ possible states. Applying any operation on that register 
afterward will be applied on the $2^3$ states simultaneously.

Controlled operations play an important role in building up quantum circuits for any given operation \cite{elementary-gates}. 
The Controlled-$U$ gate is a general controlled gate with one or more control qubit(s) 
as shown in Fig.~\ref{ENhfig3_4_X}.a. It works as follows: $U$ is applied on the target qubit 
$\left| t \right\rangle$ if and only if all $\left| {x_k } \right\rangle $ are set to $\left| 1 \right\rangle $, i.e. 
qubits will be transformed as follows,

\begin{equation}
\label{ENheq7}
\begin{array}{l}
 \left| {x_k } \right\rangle \to \left| {x_k } \right\rangle , \,k:0 \to n-1,\\ 
 \left| t \right\rangle \,\,\,\,\to  \left| {t_{CU} } \right\rangle= U^{x_0 x_1 ...x_{n-1} }\left| t \right\rangle,\\ 
 \end{array}
\end{equation}

\noindent
where $x_0 x_1 ...x_{n-1}$ in the exponent of $U$ denotes the $AND$-ing operation of 
the qubit-values $x_0,\,x_1 ,...,x_{n-1}$.

If $U$ in the general case is replaced with the $NOT$ gate mentioned above, 
the resulting gate is  called $CNOT$ gate (shown in Fig.~\ref{ENhfig3_4_X}.b). 
It inverts the target qubit if and only if all the control qubits are set to $\left| 1 \right\rangle$ as follows,

\begin{equation}
\label{ENheq8}
\begin{array}{l}
 \left| {x_k } \right\rangle \to \left| {x_k } \right\rangle ; \,k:0 \to n-1,\\ 
 \left| {t } \right\rangle \,\,\,\, \to \left| {t_{CN} } \right\rangle =  \left| {t \oplus x_0 x_2 ...x_{n-1} } 
\right\rangle, \\ 
 \end{array}
\end{equation}

\noindent
where $ \oplus $ is the classical $XOR$ operation.

\begin{center}
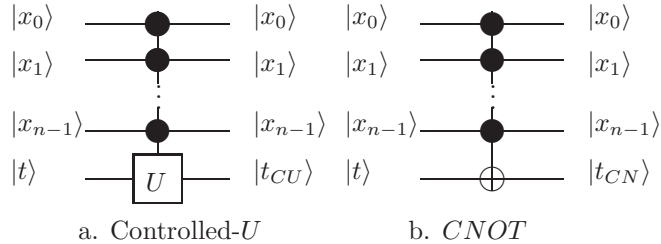
\begin{figure} [t]
\begin{center}
\setlength{\unitlength}{3947sp}%
\begingroup\makeatletter\ifx\SetFigFont\undefined%
\gdef\SetFigFont#1#2#3#4#5{%
  \reset@font\fontsize{#1}{#2pt}%
  \fontfamily{#3}\fontseries{#4}\fontshape{#5}%
  \selectfont}%
\fi\endgroup%
\begin{picture}(3525,1485)(3376,-1936)
\thinlines
\put(4201,-1261){\circle*{150}}
\put(4201,-586){\circle*{150}}
\put(4201,-811){\circle*{150}}
\put(6301,-586){\circle*{150}}
\put(6301,-811){\circle*{150}}
\put(6301,-1261){\circle*{150}}
\put(6301,-1561){\circle{150}}
\put(4201,-811){\circle*{150}}
\put(4201,-586){\circle*{150}}
\put(4201,-586){\circle{150}}
\put(4051,-1711){\framebox(300,300){}}
\put(4351,-1561){\line( 1, 0){300}}
\put(4051,-1561){\line(-1, 0){300}}
\put(4201,-1111){\line( 0,-1){300}}
\put(4651,-586){\line(-1, 0){900}}
\put(4201,-511){\line( 0,-1){450}}
\put(4651,-811){\line(-1, 0){900}}
\put(4651,-1261){\line(-1, 0){900}}
\put(6751,-586){\line(-1, 0){900}}
\put(6751,-811){\line(-1, 0){900}}
\put(6751,-1261){\line(-1, 0){900}}
\put(6751,-1561){\line(-1, 0){900}}
\put(6301,-511){\line( 0,-1){450}}
\put(6301,-1111){\line( 0,-1){525}}
\put(4185,-1120){$\vdots$}
\put(6285,-1120){$\vdots$}
\put(4801,-586){$\left| {x_0 } \right\rangle$}%
\put(3276,-586){$\left| {x_0 } \right\rangle$}%
\put(5376,-586){$\left| {x_0 } \right\rangle$}%
\put(6901,-586){$\left| {x_0 } \right\rangle$}%
\put(4801,-886){$\left| {x_1 } \right\rangle$}%
\put(3276,-886){$\left| {x_1 } \right\rangle$}%
\put(5376,-886){$\left| {x_1 } \right\rangle$}%
\put(6901,-886){$\left| {x_1 } \right\rangle$}%
\put(4801,-1561){$\left| {t_{CU} } \right\rangle$}%
\put(3276,-1561){$\left| {t } \right\rangle$}%
\put(5376,-1561){$\left| {t } \right\rangle$}%
\put(6901,-1561){$\left| {t_{CN}}  \right\rangle$}%
\put(4801,-1261){$\left| {x_{n-1} } \right\rangle$}%
\put(3276,-1261){$\left| {x_{n-1} } \right\rangle$}%
\put(5376,-1261){$\left| {x_{n-1} } \right\rangle$}%
\put(6901,-1261){$\left| {x_{n-1} } \right\rangle$}%
\put(4126,-1636){$U$}%
\put(5776,-1936){b. $CNOT$}%
\put(3700,-1936){a. Controlled-$U$}%
\end{picture}
\end{center}
\caption{Controlled gates. The back circle $\bullet $ indicates 
the control qubits, and the symbol $ \oplus $ in part (b.) indicates the target qubit.}
\label{ENhfig3_4_X}
\end{figure}
\end{center}

\subsection{Entangled States}

A state of a quantum system of two or more qubits can be represented in terms of the tensor product of each qubit. 
Sometimes it is not possible to represent the state of the system in terms of the states of its individual 
qubits. In such a case, we say that there is a correlation between these components, i.e. each component does not have 
its own state. This is usually referred to as an {\it entangled state} \cite{Bell64,Bell66,EPR35,Rieffel00}. 

For example \cite{Rieffel00}, the state $a\left| {00} \right\rangle  + b\left| {11} \right\rangle$ cannot be decomposed into the 
states of two separate qubits, i.e. we cannot find $a_0,a_1,b_0$ and $b_1$ such that,
\begin{equation}
\left( {a_0 \left| 0 \right\rangle  + b_0 \left| 1 \right\rangle } \right) \otimes \left( {a_1 \left| 0 \right\rangle  + b_1 \left| 1 \right\rangle } \right) = a\left| {00} \right\rangle  + b\left| {11} \right\rangle.
\label{ch2eqn2.15}
\end{equation} 

Entangled states are considered as the heart for many quantum algorithms, 
for example, quantum teleportation \cite{Bennett93}, dense coding \cite{Barenco95} and quantum searching \cite{Azuma00,sam02}. 
Two-qubit entangled states (shown in Eqn. \ref{bellstates}) are usually referred to as 
{\it Bell states, EPR states, EPR pairs} \cite{nc00a} or {\it Bell basis} \cite{Gruska99}.

\begin{equation}
\label{bellstates}
\frac{{\left( {\left| {00} \right\rangle  \pm \left| {11} \right\rangle } \right)}}{{\sqrt 2 }},\,\,\frac{{\left( {\left| {01} \right\rangle  \pm \left| {10} \right\rangle } \right)}}{{\sqrt 2 }}.
\end{equation}
 
\section{Basic Operations}

Before defining the operators of the QQL, three basic operations must be defined. Firstly, 
a simple way to convert the standard irreversible logic operations, 
e.g. AND, OR, NOT...etc\cite{bennett73}, to reversible logic operations 
suitable for quantum computers. This has a special importance 
in applying an arbitrary operation based on two or more SELECT operators.
Then, a quantum oracle that applies a {\it query} on a database file exists in a superposition 
and returns the result(s) of the query entangled with a temporary qubit dedicated 
for subspace identification purposes. 
Finally, an operator that acts only on a certain subspace of the system to be used 
in the process of backing up and restoring a portion of a quantum database.

\subsection{Boolean Quantum Logic ($CNOT$ gates)}

A logical expression is an expression that has two operands connected with a logical operator 
from the set $\{  > , \ge , < , \le , = , \ne \}$. 
A logical expression evaluates either to {\it true} (1) or to {\it false} (0).
A relational expression is an expression that combines two or more logical expressions with 
relational operators such as $AND$, $OR$ and $NOT$, e.g. $(x_0\,OR\,(NOT\,x_1))$, where $x_0, x_1 \in \{ 0,1\}$. 
These sort of relational expressions cannot be used directly as quantum relational expressions 
because thier operations are not reversible \cite{toffoli80}. 
A relational expression can be understood as a Boolean function while the logical expressions are 
the Boolean inputs to that Boolean function .  

In building quantum circuits for Boolean functions, an extra temporary qubit will be added to the system and 
will be initialized to state $\left| 0 \right\rangle$, to hold the result of the Boolean function at the end of the computation. 
For clarity purposes, the $CNOT$ gates will be presented as follows \cite{transrules}: $CNOT(C \vert t)$ is a gate where the target qubit $\left| t \right\rangle$ is 
controlled by a set of qubits $C$ such that $t \notin C$, the state of the 
qubit $\left| t \right\rangle$ will be flipped from $\left| 0 \right\rangle $ to $\left| 1 
\right\rangle $ or from $\left| 1 \right\rangle $ to $\left| 0 \right\rangle 
$ if and only if all the qubits in $C$ are set to true (state $\left| 1 
\right\rangle )$, i.e. the new state of the target qubit $\left| t \right\rangle$ will be the result 
of $XOR$-ing the old state of $\left| t \right\rangle$ with the $AND$-ing of the states of the control 
qubits. For example, consider the $CNOT$ gate shown in Fig.~\ref{fig1}, it can be represented 
as $CNOT\left( {\left\{ {x_0 ,x_2 } 
\right\}\vert x_3 } \right)$, where $\bullet $ on a qubit means that the 
condition on that qubit will evaluate to true if and only if the state of 
that qubit is $\left| 1 \right\rangle $, while $ \oplus $ denotes the target 
qubit which will be flipped if and only if all the control qubits are set 
to true, which means that the state of the qubit $\left| x_3 \right\rangle$ will be flipped if 
and only if $\left|x_{0}\right\rangle=\left|x_{2}\right\rangle=\left| 1 \right\rangle $ with whatever value 
in $\left|x_{1}\right\rangle$; i.e. $\left|x_{3}\right\rangle$ will be changed according to the operation $x_3 \to 
x_3 \oplus x_0 x_2 $. If $C=\Phi$, i.e. an empty set, then the target qubit will be flipped 
unconditionally ($NOT$ gate).

\begin{center}
\begin{figure}  [t]
\begin{center}
\setlength{\unitlength}{3947sp}%
\begingroup\makeatletter\ifx\SetFigFont\undefined%
\gdef\SetFigFont#1#2#3#4#5{%
  \reset@font\fontsize{#1}{#2pt}%
  \fontfamily{#3}\fontseries{#4}\fontshape{#5}%
  \selectfont}%
\fi\endgroup%
\begin{picture}(624,1070)(2389,-2548)
\thinlines
\put(2701,-1561){\circle*{150}}
\put(2701,-2161){\circle*{150}}
\put(2701,-2461){\circle{150}}
\put(2401,-1561){\line( 1, 0){600}}
\put(2100,-1561){$\left| {x_0 } \right\rangle$}
\put(2401,-1861){\line( 1, 0){ 75}}
\put(2100,-1861){$\left| {x_1 } \right\rangle$}
\put(2476,-1861){\line( 1, 0){ 75}}
\put(2551,-1861){\line( 1, 0){ 75}}
\put(2626,-1861){\line( 1, 0){ 75}}
\put(2701,-1861){\line( 1, 0){ 75}}
\put(2776,-1861){\line( 1, 0){ 75}}
\put(2851,-1861){\line( 1, 0){ 75}}
\put(2926,-1861){\line( 1, 0){ 75}}
\put(2401,-2161){\line( 1, 0){ 75}}
\put(2100,-2161){$\left| {x_2 } \right\rangle$}
\put(2476,-2161){\line( 1, 0){ 75}}
\put(2551,-2161){\line( 1, 0){ 75}}
\put(2626,-2161){\line( 1, 0){ 75}}
\put(2701,-2161){\line( 1, 0){ 75}}
\put(2776,-2161){\line( 1, 0){ 75}}
\put(2851,-2161){\line( 1, 0){ 75}}
\put(2926,-2161){\line( 1, 0){ 75}}
\put(2401,-2461){\line( 1, 0){600}}
\put(2100,-2461){$\left| {x_3 } \right\rangle$}
\put(2701,-1561){\line( 0,-1){975}}
\end{picture}
\end{center}
\caption{\label{fig1}$CNOT\left( {\left\{ {x_0 ,x_2 } \right\}\vert x_3 } \right)$ gate.}
\end{figure}
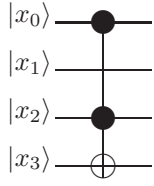
\end{center}

\subsection{Boolean Quantum Circuits (BQC)}

A general Boolean quantum circuit $U$ of size $m$ (size of the circuit refers to the total number of 
$CNOT$ gates in that circuit) over $n$ qubit quantum system 
with qubits $\left| {x_0 } \right\rangle ,\left| {x_1 } \right\rangle 
,\ldots ,\left| {x_{n - 1} } \right\rangle $ can be represented as a 
sequence of $CNOT$ gates \cite{transrules} as follows,

\begin{equation}
\label{eqn5}
U_g = CNOT\left( {C_1 \vert t_1 } \right)\ldots CNOT\left( {C_j \vert t_j } 
\right)\ldots CNOT\left( {C_m \vert t_m } \right),
\end{equation}

\noindent
where $t_j \in \left\{ {x_0 ,\ldots ,x_{n - 1} } \right\}$; $C_j \subset 
\left\{ {x_0 ,\ldots ,x_{n - 1} } \right\}$; $t_j \notin C_j$ and $j:1\to m$. 
The BQC that will be used in this paper can be represented as follows,

\begin{equation}
\label{eqn6}
{U} = CNOT(C_1 \vert t)...CNOT(C_j \vert t)...CNOT(C_m \vert t),
\end{equation}

\noindent
where $t \equiv x_{n - 1} ;\,\,C_j \subseteq \left\{ {x_0 ,\ldots ,x_{n - 2} } 
\right\}$. For example, consider the quantum circuit shown in Fig.~\ref{fig3}, it can be 
represented as follows,

\begin{equation}
\label{eqn7}
U = CNOT(\{x_0 ,x_1 \}\vert x_2 ).CNOT(\{x_1 \}\vert x_2 ).CNOT(x_2 ),
\end{equation}

\begin{center}
\begin{figure} [t]
\begin{center}
\setlength{\unitlength}{3947sp}%
\begingroup\makeatletter\ifx\SetFigFont\undefined%
\gdef\SetFigFont#1#2#3#4#5{%
  \reset@font\fontsize{#1}{#2pt}%
  \fontfamily{#3}\fontseries{#4}\fontshape{#5}%
  \selectfont}%
\fi\endgroup%
\begin{picture}(1362,792)(1051,-2248)
\thinlines
\put(1501,-1561){\circle*{150}}
\put(1501,-1861){\circle*{150}}
\put(1801,-1861){\circle*{150}}
\put(2101,-2161){\circle{150}}
\put(1801,-2161){\circle{150}}
\put(1501,-2161){\circle{150}}
\put(1201,-1561){\line( 1, 0){1200}}
\put(1201,-1861){\line( 1, 0){1200}}
\put(1201,-2161){\line( 1, 0){1200}}
\put(1501,-2236){\line( 0, 1){750}}
\put(1501,-1486){\line( 0,-1){ 75}}
\put(1801,-2236){\line( 0, 1){375}}
\put(2101,-2236){\line( 0, 1){150}}
\put(920,-1561){$\left| {x_0 } \right\rangle$}%
\put(920,-1861){$\left| {x_1 } \right\rangle$}%
\put(920,-2161){$\left| {x_2 } \right\rangle$}%
\end{picture}
\end{center}
\caption{\label{fig3}Boolean quantum circuit.}
\end{figure}
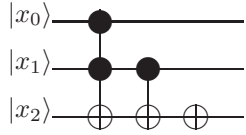
\end{center}

Now, to trace the operations that have been applied on the target qubit $\left|x_{2}\right\rangle$, we 
will trace the operation of each of the $CNOT$ gates that has been applied:

\begin{itemize}
\item $CNOT(\{x_0 ,x_1 \}\vert x_2 ) \Rightarrow x_2 \to x_2 \oplus x_0 x_1$,
\item $CNOT(\{x_1 \}\vert x_2 ) \Rightarrow x_2 \to x_2 \oplus x_1$ ,
\item $CNOT(x_2 ) \Rightarrow x_2 \to \overline x _2 = x_2 \oplus 1$.
\end{itemize}

Combining the three operations, we see that the complete operation applied on 
$\left|x_{2}\right\rangle$ is represented as follows,

\begin{equation}
\label{eqn8}
x_2 \to x_2 \oplus x_0 x_1 \oplus x_1 \oplus 1.
\end{equation}

If $\left|x_{2}\right\rangle$ is initialized to $\left| 0 \right\rangle $, applying the 
circuit will make $\left|x_{2}\right\rangle$ carry the result of the operation ($x_0 x_1 \oplus 
x_1 \oplus 1)$, which is equivalent to the operation $x_0 + \overline x _1 $, i.e.  $(x_0\,OR\,(NOT\,x_1))$. 
More detials on how to convert more complex {\it canonical Boolean expression} 
(expressions use $AND$, $OR$, $NOT$) to quantum circuits using {\it Reed-Muller expression} (expressions use $AND$, $XOR$, $NOT$) 
can be found in \cite{Younes03b}.

\subsection{Quantum Oracle}

Consider an unstructured list $L$ of $N$ items. For simplicity and without loss of generality we will 
assume that $N = 2^n$ for some positive integer $n$. Suppose the items in the list are labeled with the 
integers $\{0,1,...,N - 1\}$, and consider a Boolean function $f$ which maps an item $i \in L$ to 
either 0 or 1 according to some properties this item should satisfy, i.e. $f:L \to \{ 0,1\}$. 

It follows directly, from the discussion in the above sections, that the function $f$ can be represented 
as a unitary matrix $U_f$. 
$U_f$ will be taken as an oracle that applies a {\it query} on the database file and returns the results. 
$U_f$ has the following effect when applied on a quantum register $\left| x,y \right\rangle$,

\begin{equation}
U_f :\left| {x,y} \right\rangle  \to \left| {x,y \oplus f(x)} \right\rangle, 
\end{equation}

\noindent
where $\left| x \right\rangle$ is a quantum register of size $n$ and $\left| y \right\rangle$ is 
a temporary qubit. If $\left| y \right\rangle$ is initially set to $\left| 0 \right\rangle$, then $U_f$ has the following 
effect on the quantum register,

\begin{equation}
U_f :\left| {x,0} \right\rangle  \to \left| {x,f(x)} \right\rangle. 
\end{equation}

This oracle has a special importance in setting up an entanglement on the states that make the 
oracle evaluates to true as follows: assume that $\left| \psi  \right\rangle$ is a quantum register 
of size $n+1$ qubits. The first $n$ qubits in a superposition and the last qubit is an extra qubit initialized 
to state $\left| 0  \right\rangle$. Assume that $U_f$ is a quantum oracle used to identify 
the states in the superposition that make $f$ evaluates to true. Applying $U_f$ on $\left| \psi  \right\rangle$ 
can be understood as follows,

\begin{equation}
\begin{array}{l}
 U_f \left| \psi  \right\rangle  = U_f \sum\limits_{i = 0}^{2^n  - 1} {\alpha _i \left| i \right\rangle  \otimes \left| 0 \right\rangle  = } \sum\limits_{i = 0}^{2^n  - 1} {\alpha _i \left| i \right\rangle  \otimes \left| {f(i)} \right\rangle }  \\ 
 \,\,\,\,\,\,\,\,\,\,\,\,\,\,\,\,\,\, = \sum\limits_{i = 0}^{2^n  - 1} {^{'} \alpha _i \left| i \right\rangle  \otimes \left| 1 \right\rangle  + } \sum\limits_{i = 0}^{2^n  - 1} {^{''} \alpha _i \left| i \right\rangle  \otimes \left| 0 \right\rangle },  \\ 
 \end{array}
\end{equation}

\noindent
where, $\sum\nolimits_i {^{'}} $ denotes a sum over $i$ which are desired items, 
and $\sum\nolimits_i {^{''}} $ denotes a sum over $i$ which are undesired items in the list, i.e. 
the list of desired items are entangled with state $\left| 1  \right\rangle$ of the extra qubit and 
the list of undesired items are entangled with state $\left| 0  \right\rangle$. So far, this can 
be considered as the SELECT operator since the selected states is entangled with state $\left| 1  \right\rangle$.
Applying any operation $U$ based on the condition that the extra qubit is in state $\left| 1  \right\rangle$ 
will be applied only of the subspace of the desired items as shown in Fig.~\ref{figQO}. To apply an arbitrary operation $U$ ($2^n\times2^n$ unitary matrix) 
only on the subspace entangled with state $\left| 1  \right\rangle$, $U$ must be transformed to 
a unitary matrix of size $2^{n+1} \times 2^{n+1}$ as follows,

\begin{equation}
U \to U \otimes \left| 1 \right\rangle \left\langle 1 \right| + I_n  \otimes \left| 0 \right\rangle \left\langle 0 \right|,
\label{ContU}
\end{equation}

\noindent
where $I_n$ is the identity matrix of size $2^n\times 2^n$.

\begin{center}
\begin{figure} [t]
\begin{center}
\setlength{\unitlength}{3947sp}%
\begingroup\makeatletter\ifx\SetFigFont\undefined%
\gdef\SetFigFont#1#2#3#4#5{%
  \reset@font\fontsize{#1}{#2pt}%
  \fontfamily{#3}\fontseries{#4}\fontshape{#5}%
  \selectfont}%
\fi\endgroup%
\begin{picture}(3489,1922)(1580,-2406)
{\thinlines
\put(2921,-2229){\circle{134}}
}%
{\put(4241,-2234){\circle*{134}}
}%
{\put(3721,-1842){\framebox(1013,1325){}}
}%
{\put(2119,-580){\line( 1, 0){296}}
}%
{\put(2126,-813){\line( 1, 0){296}}
}%
{\put(2426,-1836){\framebox(1013,1325){}}
}%
{\put(2126,-1759){\line( 1, 0){296}}
}%
{\put(4741,-579){\line( 1, 0){296}}
}%
{\put(4744,-815){\line( 1, 0){296}}
}%
{\put(4736,-1759){\line( 1, 0){296}}
}%

{\put(3440,-576){\line( 1, 0){271}}}

{\put(3448,-811){\line( 1, 0){271}}}

{\put(3448,-1763){\line( 1, 0){271}}}

{\put(2125,-2229){\line( 1, 0){2932}}
}%
{\put(2918,-2289){\line( 0, 1){458}}
}%
{\put(2169,-496){\line(-1, 0){116}}
}%
{\put(2053,-496){\line( 0,-1){1331}}
\put(2053,-1827){\line( 1, 0){120}}
}%
{\put(4242,-2168){\line( 0, 1){326}}
}%

\put(2176,-1386){$\vdots$}

\put(4876,-1386){$\vdots$}

\put(4177,-1225){\makebox(0,0)[lb]{\smash{{\SetFigFont{12}{14.4}{\rmdefault}{\mddefault}{\updefault}{$U$}%
}}}}
\put(2854,-1217){\makebox(0,0)[lb]{\smash{{\SetFigFont{12}{14.4}{\rmdefault}{\mddefault}{\updefault}{$U_f$}%
}}}}
\put(1658,-966){\makebox(0,0)[lb]{\smash{{\SetFigFont{12}{14.4}{\rmdefault}{\mddefault}{\updefault}{$n$}%
}}}}
\put(1495,-1125){\makebox(0,0)[lb]{\smash{{\SetFigFont{12}{14.4}{\rmdefault}{\mddefault}{\updefault}{qubits}%
}}}}
\put(1638,-2377){\makebox(0,0)[lb]{\smash{{\SetFigFont{12}{14.4}{\rmdefault}{\mddefault}{\updefault}{qubit}%
}}}}
\put(1654,-2181){\makebox(0,0)[lb]{\smash{{\SetFigFont{12}{14.4}{\rmdefault}{\mddefault}{\updefault}{extra}%
}}}}
\end{picture}%

\end{center}
\caption{\label{figQO}Setting up entanglement on a subspace of the superposition.}
\end{figure}
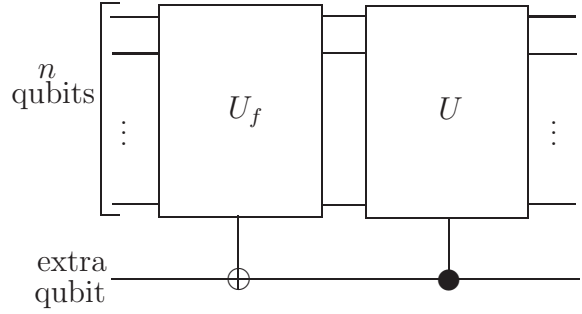
\end{center}

\subsection{Partial Diffusion}

The {\it partial diffusion operator}, $D_p$, is an operator that performs amplitude alteration only 
on the subspace of the system entangled with the extra qubit workspace in state $\left|0\right\rangle$ \cite{Younes03d}. 
The diagonal representation of $D_p$ when applied on $n+1$ qubits system can take this form:

\begin{equation}
\label{ENheq13}
D_p =  \left(H^{ \otimes n}  \otimes I_1\right)\left( {(1 - e^{i\varphi } )\left| 0 \right\rangle \left\langle 0 \right| - I_{n+1}} \right)\left(H^{ \otimes n}  \otimes I_1\right),
\end{equation}
\noindent
where the vector $\left| 0 \right\rangle$ used in Eqn. \ref{ENheq13} is of length $2N=2^{n+1}$, $I_k$ is the identity matrix 
of size $2^k\times 2^k$ and $\varphi$ is an arbitrary angle. Consider a general state $\left|\psi\right\rangle$ of $n+1$ qubits register:

\begin{equation}
\begin{array}{l}
\left| \psi  \right\rangle  = \sum\limits_{k = 0}^{2N - 1} {\delta _k \left| k \right\rangle } = \sum\limits_{j = 0}^{N - 1} {\alpha _j \left( {\left| j \right\rangle  \otimes \left| 0 \right\rangle } \right)}  + \sum\limits_{j = 0}^{N - 1} {\beta _j \left( {\left| j \right\rangle  \otimes \left| 1 \right\rangle } \right)},
\end{array}
\end{equation}

\noindent
where \{$\alpha _j  = \delta _k$ : $k$ even\} and \{$\beta _j  = \delta _k$ : $k$ odd\}. 
The effect of applying $D_p$ on $\left| \psi  \right\rangle$ produces,

\begin{equation}
\label{ENheq15}
\begin{array}{l}
D_p\left| \psi  \right\rangle = \left( H^{ \otimes n}  \otimes I_1 \right) \left( {(1 - e^{i\varphi } )\left| 0 \right\rangle \left\langle 0 \right| - I_{n+1}} \right) \left( H^{ \otimes n}  \otimes I_1 \right)\sum\limits_{k = 0}^{2N - 1} {\delta _k \left| k \right\rangle }\\
\,\,\,\,\,\,\,\,\,\,\,\,\,\,\,\,\,\, =\sum\limits_{j = 0}^{N - 1} {(1 - e^{i\varphi } )\left\langle \alpha  \right\rangle \left( {\left| j \right\rangle  \otimes \left| 0 \right\rangle } \right)}  - \sum\limits_{k = 0}^{2N - 1} {\delta _k } \left| k \right\rangle\\ 
\,\,\,\,\,\,\,\,\,\,\,\,\,\,\,\,\,\,=\sum\limits_{j = 0}^{N - 1} {\left( {(1 - e^{i\varphi } )\left\langle \alpha  \right\rangle  - \alpha _j } \right)\left( {\left| j \right\rangle  \otimes \left| 0 \right\rangle } \right)}  - \sum\limits_{j = 0}^{N - 1} {\beta _j \left( {\left| j \right\rangle  \otimes \left| 1 \right\rangle } \right)},\\
\end{array}
\end{equation}

\noindent
where $\left\langle \alpha  \right\rangle  = \frac{1}{N}\sum\nolimits_{j = 0}^{N - 1} {\alpha _j }$ is the mean of 
the amplitudes of the subspace ${\alpha _j \left( {\left| j \right\rangle  \otimes 
\left| 0 \right\rangle } \right)}$, i.e. applying the operator $D_p$ will only alter the amplitudes of the 
subspace ${\alpha _j \left( {\left| j \right\rangle  \otimes \left| 0 \right\rangle } \right)}$ 
and will only {\it change the sign} of the amplitudes for the subspace ${\beta _j \left( {\left| j \right\rangle  \otimes \left| 1 \right\rangle } \right)}$.
If $\varphi=\pi$, $D_p$ will perform the inversion about the mean only on the subspace ${\alpha _j \left( {\left| j \right\rangle  \otimes \left| 0 \right\rangle } \right)}$ \cite{Younes03d}. 
For simplicity and without loss of generality, we will use $D_p$ with $\varphi=\pi$ throughout the rest of the paper.

\section{Quantum Query Language}

The architecture of the memory of the quantum system required for the operations of the QQL  
consists of a quantum register of size $n+t$ qubits. Initially, the system is set to  
state $\left| 0 \right\rangle ^{ \otimes n}  \otimes \left| 0 \right\rangle ^{ \otimes t}$. 
The $n$ qubits can hold up to $2^n$ records at a time and the $t$ qubits will be used 
as temporary qubits for processing purposes. If it is required to store $r$ records in 
a superposition such that $1 \le r \le 2^n$, then $\left\lceil {\log _2 (r)} \right\rceil$ qubits 
will be used out of the $n$ qubits. 

It is important to clearly declare that the following QQL operators care only about the effects
 to be applied on the states of the system (values in the list). For simplicity, the 
 effects to be applied on the amplitudes associated 
with the states in the superposition have been ignored as long as the required states 
exist in the superposition. The QQL operators could be associated 
with some quantum operators, to be constructed separately, 
for amplitude manipulation and to maintain the stability of the amplitudes during the processing time in specific situations.

\subsection{Inserting Records to the Superposition (INSERT)}

Suppose that it is required to insert some records to a superposition. 
To insert $2^r$ records directly to the superposition such that 
$r \le n$, apply $H^{ \otimes r}  \otimes I^{ \otimes n - r}$ on the first $r$ qubits to create 
a system in a superposition as follows,

\begin{equation}
\left( {\sum\limits_{i = 0}^{2^r  - 1} {\alpha _i \left| i \right\rangle } } \right) \otimes \left| 0 \right\rangle ^{ \otimes n - r}.
\end{equation}

If it is required to insert certain number of records $r$ to a superposition such that only one record 
is inserted at a time, then controlled 
Hadamard gates can be used to achieve this goal. For example, assume that 
there is a quantum register of three qubits that can hold up to eight values. To insert item-by-item 
in sequence to the superposition, apply in sequence the set of operators 
$S_i ,i = 0,\ldots,7$ defined as follows (as shown in Fig.~\ref{insertfig}),

\begin{equation}
\begin{array}{l}
 S_1  = H \otimes I \otimes I, \\ 
 S_2  = \left| 0 \right\rangle \left\langle 0 \right| \otimes H \otimes I + \left| 1 \right\rangle \left\langle 1 \right| \otimes I \otimes I,\\
 S_3  = \left| 0 \right\rangle \left\langle 0 \right| \otimes I \otimes I + \left| 1 \right\rangle \left\langle 1 \right| \otimes H \otimes I,\\ 
 S_4  = \left| {00} \right\rangle \left\langle {00} \right| \otimes H + \sum\limits_{i = 0,i \ne 0}^3 {\left| i \right\rangle \left\langle i \right|}  \otimes I,\\
 S_5  = \left| {10} \right\rangle \left\langle {10} \right| \otimes H + \sum\limits_{i = 0,i \ne 2}^3 {\left| i \right\rangle \left\langle i \right|}  \otimes I, \\ 
 S_6  = \left| {01} \right\rangle \left\langle {01} \right| \otimes H + \sum\limits_{i = 0,i \ne 1}^3 {\left| i \right\rangle \left\langle i \right|}  \otimes I,\\ 
 S_7  = \left| {11} \right\rangle \left\langle {11} \right| \otimes H + \sum\limits_{i = 0,i \ne 3}^3 {\left| i \right\rangle \left\langle i \right|}  \otimes I. \\ 
 \end{array}
\end{equation}

Initially, the system is in state $\left| {000} \right\rangle$, so, the system 
already contains an item. To insert the $2^{nd}$ item, apply $S_1$, so the system is transoformed 
to the following,

\begin{equation}
\alpha _0 \left| {000} \right\rangle  + \alpha _1 \left| {001} \right\rangle,
\end{equation}
 
\noindent
and, to insert the $3^{rd}$ item, apply $S_2$ to get,

\begin{equation}
\alpha _{00} \left| {000} \right\rangle  + \alpha _{01} \left| {001} \right\rangle  + \alpha _{10} \left| {010} \right\rangle, 
\end{equation}

\noindent
and so on. If we keep applying $S_i's$ up to $S_6$, we get,

\begin{equation}
\alpha _{000} \left| {000} \right\rangle  + \alpha _{001} \left| {001} \right\rangle  + \alpha _{010} \left| {010} \right\rangle  + \alpha _{011} \left| {011} \right\rangle  + \alpha _{100} \left| {100} \right\rangle  + \alpha _{101} \left| {101} \right\rangle  + \alpha _{110} \left| {110} \right\rangle.
\end{equation}

Finally, applying $S_7$ will {\it complete} the superposition over the whole quantum register. To speed up 
this process a little bit, assume that it is required to insert five records to the superposition, 
then, firstly, apply $H \otimes H \otimes I$, to insert four records directly to the superposition in a single step, 
since $H \otimes H \otimes I = S_3 S_2 S_1$, then apply $S_4$ to insert the $5^{th}$ record. The natural question that might arise here is: 
What if it is required to insert some specific states, not necessarily in sequence, to the superposition? 
The answer might be more obvious after the UPDATE operator is defined in the next section.

\begin{center}
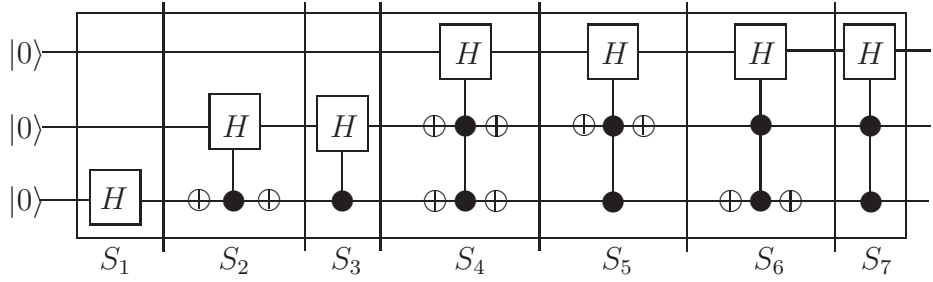
\begin{figure}  [htbp]
\begin{center}

\setlength{\unitlength}{3947sp}%
\begingroup\makeatletter\ifx\SetFigFont\undefined%
\gdef\SetFigFont#1#2#3#4#5{%
  \reset@font\fontsize{#1}{#2pt}%
  \fontfamily{#3}\fontseries{#4}\fontshape{#5}%
  \selectfont}%
\fi\endgroup%
\begin{picture}(5836,2045)(242,-6324)
{\put(2373,-5537){\circle*{130}}
}%
{\put(1912,-5532){\circle{130}}
}%
{\put(3341,-5070){\circle{130}}
}%
{\put(3336,-5537){\circle{130}}
}%
{\put(4074,-5067){\circle*{130}}
}%
{\put(3889,-5064){\circle{130}}
}%
{\put(4071,-5543){\circle*{130}}
}%
{\put(4263,-5068){\circle{130}}
}%
{\put(3144,-5070){\circle*{130}}
}%
{\put(3144,-5542){\circle*{130}}
}%
{\put(2954,-5539){\circle{130}}
}%
{\put(2951,-5070){\circle{130}}
}%
{\put(1690,-5542){\circle*{130}}
}%
{\put(1472,-5534){\circle{130}}
}%
{\put(5000,-5542){\circle*{130}}
}%
{\put(4810,-5539){\circle{130}}
}%
{\put(5188,-5542){\circle{130}}
}%
{\put(5001,-5066){\circle*{130}}
}%
{\put(5688,-5071){\circle*{130}}
}%
{\put(5692,-5545){\circle*{130}}
}%
{\put(789,-5686){\framebox(312,338){}}
}%
{\put(1539,-5211){\framebox(312,338){}}
}%
{\put(2214,-5224){\framebox(312,338){}}
}%
{\put(2989,-4774){\framebox(312,338){}}
}%
{\put(3914,-4774){\framebox(312,338){}}
}%
{\put(4839,-4774){\framebox(312,338){}}
}%
{\put(5527,-4774){\framebox(312,338){}}
}%
{\put(707,-5772){\framebox(5200,1411){}}
}%
{\put(1109,-5540){\line( 1, 0){4944}}
}%
{\put(488,-5074){\line( 1, 0){1056}}
}%
{\put(1857,-5067){\line( 1, 0){360}}
}%
{\put(2534,-5067){\line( 1, 0){3528}}
}%
{\put(488,-4601){\line( 1, 0){2496}}
}%
{\put(3310,-4601){\line( 1, 0){600}}
}%
{\put(4235,-4601){\line( 1, 0){600}}
}%
{\put(5850,-4594){\line( 1, 0){216}}
}%
{\put(5159,-4594){\line( 1, 0){360}}
}%
{\put(2135,-4291){\line( 0,-1){1560}}
}%
{\put(2609,-4291){\line( 0,-1){1560}}
}%
{\put(3610,-4291){\line( 0,-1){1560}}
}%
{\put(4534,-4298){\line( 0,-1){1560}}
}%
{\put(5465,-4298){\line( 0,-1){1560}}
}%
{\put(1247,-4297){\line( 0,-1){1560}}
}%
{\put(470,-5533){\line( 1, 0){312}}
}%
{\put(4069,-4777){\line( 0,-1){816}}
\put(4069,-5593){\line( 0, 1){  0}}
}%
{\put(3141,-4773){\line( 0,-1){816}}
}%
{\put(2368,-5227){\line( 0,-1){360}}
\put(2368,-5587){\line( 0, 1){  0}}
}%
{\put(1686,-5214){\line( 0,-1){360}}
}%
{\put(4998,-4783){\line( 0,-1){816}}
\put(4998,-5599){\line( 0, 1){  0}}
}%
{\put(5686,-4783){\line( 0,-1){816}}
\put(5686,-5599){\line( 0, 1){  0}}
}%
{\put(1469,-5468){\line( 0,-1){120}}
}%
{\put(1910,-5468){\line( 0,-1){120}}
}%
{\put(2954,-5482){\line( 0,-1){120}}
}%
{\put(2947,-5005){\line( 0,-1){120}}
}%
{\put(3332,-5478){\line( 0,-1){120}}
}%
{\put(3339,-5008){\line( 0,-1){120}}
}%
{\put(3888,-4998){\line( 0,-1){120}}
}%
{\put(4260,-5004){\line( 0,-1){120}}
}%
{\put(4807,-5482){\line( 0,-1){120}}
}%
{\put(5185,-5479){\line( 0,-1){120}}
}%
\put(847,-5968){\makebox(0,0)[lb]{\smash{{\SetFigFont{12}{14.4}{\rmdefault}{\mddefault}{\updefault}{$S_1$}%
}}}}
\put(1592,-5968){\makebox(0,0)[lb]{\smash{{\SetFigFont{12}{14.4}{\rmdefault}{\mddefault}{\updefault}{$S_2$}%
}}}}
\put(2298,-5968){\makebox(0,0)[lb]{\smash{{\SetFigFont{12}{14.4}{\rmdefault}{\mddefault}{\updefault}{$S_3$}%
}}}}
\put(3075,-5968){\makebox(0,0)[lb]{\smash{{\SetFigFont{12}{14.4}{\rmdefault}{\mddefault}{\updefault}{$S_4$}%
}}}}
\put(3998,-5968){\makebox(0,0)[lb]{\smash{{\SetFigFont{12}{14.4}{\rmdefault}{\mddefault}{\updefault}{$S_5$}%
}}}}
\put(4953,-5968){\makebox(0,0)[lb]{\smash{{\SetFigFont{12}{14.4}{\rmdefault}{\mddefault}{\updefault}{$S_6$}%
}}}}
\put(5628,-5968){\makebox(0,0)[lb]{\smash{{\SetFigFont{12}{14.4}{\rmdefault}{\mddefault}{\updefault}{$S_7$}%
}}}}
\put(270,-4672){\makebox(0,0)[lb]{\smash{{\SetFigFont{12}{14.4}{\rmdefault}{\mddefault}{\updefault}{$\left| 0 \right\rangle$}%
}}}}
\put(270,-5145){\makebox(0,0)[lb]{\smash{{\SetFigFont{12}{14.4}{\rmdefault}{\mddefault}{\updefault}{$\left| 0 \right\rangle$}%
}}}}
\put(270,-5595){\makebox(0,0)[lb]{\smash{{\SetFigFont{12}{14.4}{\rmdefault}{\mddefault}{\updefault}{$\left| 0 \right\rangle$}%
}}}}
\put(849,-5593){\makebox(0,0)[lb]{\smash{{\SetFigFont{12}{14.4}{\rmdefault}{\mddefault}{\updefault}{$H$}%
}}}}
\put(1616,-5136){\makebox(0,0)[lb]{\smash{{\SetFigFont{12}{14.4}{\rmdefault}{\mddefault}{\updefault}{$H$}%
}}}}
\put(2294,-5136){\makebox(0,0)[lb]{\smash{{\SetFigFont{12}{14.4}{\rmdefault}{\mddefault}{\updefault}{$H$}%
}}}}
\put(3077,-4679){\makebox(0,0)[lb]{\smash{{\SetFigFont{12}{14.4}{\rmdefault}{\mddefault}{\updefault}{$H$}%
}}}}
\put(3995,-4679){\makebox(0,0)[lb]{\smash{{\SetFigFont{12}{14.4}{\rmdefault}{\mddefault}{\updefault}{$H$}%
}}}}
\put(4917,-4679){\makebox(0,0)[lb]{\smash{{\SetFigFont{12}{14.4}{\rmdefault}{\mddefault}{\updefault}{$H$}%
}}}}
\put(5600,-4679){\makebox(0,0)[lb]{\smash{{\SetFigFont{12}{14.4}{\rmdefault}{\mddefault}{\updefault}{$H$}%
}}}}

\end{picture}%

\end{center}
\caption{Sequential insertion of items to a superposition.}
\label{insertfig}
\end{figure}
\end{center}

\subsection{Updating a Set of Records (UPDATE)}

Updating a record is just sending the state that represents that record to another state that 
represents the updated record such that the record remains unique within the context 
of the database file. For example, assume that we have some records in a superposition as following,

\begin{equation}
\alpha _{000} \left| {000} \right\rangle  + \alpha _{010} \left| {010} \right\rangle  + \alpha _{011} \left| {011} \right\rangle  + \alpha _{101} \left| {101} \right\rangle  + \alpha _{110} \left| {110} \right\rangle. 
\label{b4update}
\end{equation}

To update the record $\left| {011} \right\rangle$ to be $\left| {111} \right\rangle$, i.e. it is required 
to tranform the system shown in Eqn.(\ref{b4update}) to the following system,
\begin{equation}
\alpha _{000}\left| {000} \right\rangle  + \alpha _{010} \left| {010} \right\rangle  + \alpha _{011} \left| {111} \right\rangle  + \alpha _{101} \left| {101} \right\rangle  + \alpha _{110} \left| {110} \right\rangle, 
\label{afterupdate}
\end{equation}

\noindent
such that no change in the amplitude of the updated record, then this is a permutation. 
A permutation operator is a widely known operator that can be represented as a unitary matrix 
with 0's and 1's as its entries such that each row and column contains a single 1 and 0 everywhere else. 
So, the {\it UPDATE} operator that will transform the superposition in Eqn.(\ref{b4update}) to 
the superposition in Eqn.(\ref{afterupdate}) can be written as follows,

\begin{equation}
U_{\left| {011} \right\rangle  \leftrightarrow \left| {111} \right\rangle }=\left[ {\begin{array}{*{20}c}
   1 & 0 & 0 & 0 & 0 & 0 & 0 & 0  \\
   0 & 1 & 0 & 0 & 0 & 0 & 0 & 0  \\
   0 & 0 & 1 & 0 & 0 & 0 & 0 & 0  \\
   0 & 0 & 0 & 0 & 0 & 0 & 0 & 1  \\
   0 & 0 & 0 & 0 & 1 & 0 & 0 & 0  \\
   0 & 0 & 0 & 0 & 0 & 1 & 0 & 0  \\
   0 & 0 & 0 & 0 & 0 & 0 & 1 & 0  \\
   0 & 0 & 0 & 1 & 0 & 0 & 0 & 0  \\
\end{array}} \right].
\label{update1}
\end{equation}

The UPDATE operator shown in Eqn.(\ref{update1}) is just an identity matrix of 
size $2^3 \times 2^3 $ (3-qubit register) with the $4^{th} (\left| {011} \right\rangle)$ 
and $8^{th} (\left| {111} \right\rangle)$ columns 
been {\it swapped} together to affect the basis of the system as required. Notice that, applying 
$U_{\left| {011} \right\rangle  \leftrightarrow \left| {111} \right\rangle}$ shown in Eqn.(\ref{update1}) again will {\it undo} the update.
More update operations can be achieved using a single UPDATE operator. For example, to update the 
records $\left| {000} \right\rangle$ and $\left| {010} \right\rangle$ to states $\left| {100} \right\rangle$ 
and $\left| {001} \right\rangle$ 
respectively, a single UPDATE operator is required as follows,

\begin{equation}
U_{\scriptstyle \left| {000} \right\rangle  \leftrightarrow \left| {100} \right\rangle  \hfill \atop 
  \scriptstyle \left| {010} \right\rangle  \leftrightarrow \left| {001} \right\rangle  \hfill} 
=\left[ {\begin{array}{*{20}c}
   0 & 0 & 0 & 0 & 1 & 0 & 0 & 0  \\
   0 & 0 & 1 & 0 & 0 & 0 & 0 & 0  \\
   0 & 1 & 0 & 0 & 0 & 0 & 0 & 0  \\
   0 & 0 & 0 & 1 & 0 & 0 & 0 & 0  \\
   1 & 0 & 0 & 0 & 0 & 0 & 0 & 0  \\
   0 & 0 & 0 & 0 & 0 & 1 & 0 & 0  \\
   0 & 0 & 0 & 0 & 0 & 0 & 1 & 0  \\
   0 & 0 & 0 & 0 & 0 & 0 & 0 & 1  \\
\end{array}} \right].
\end{equation}

A quantum circuit can be constructed for such permutation matrices using elementary $CNOT$ gates \cite{Younes03a}. 
We may conclude from the INSERT and UPDATE operators that any arbitrary records can be included in 
a superposition. They are not necessarily to be in sequence. This can be done 
by inserting the required number of states, then apply an UPDATE operator on some states 
to get the final required states in the superposition.   
  
\subsection{Deleting a Set of Records (DELETE)}

Assume that we want to delete some specific records from the superposition. This problem is an interesting 
problem by itself. How can we remove some items from a superposition in a single step? The 
answer to this question is still quite open. In this section, we will discuss some key points that might be used to solve this 
problem. Firstly, we need to identify the items to be removed from the superposition. Assume that 
we have a Boolean function $f$ that evaluates to true for the items we want to delete. Applying a quantum oracle $U_f$ 
on the superposition taking a temporary qubit as the target qubit will identify these items by entangling 
the subspace of the items we want to keep in the superposition with state 
$\left| 0 \right\rangle$ of the temporary qubit, and the subspace of the items we want to delete 
with state $\left| 1 \right\rangle$ of the temporary qubit. The rest is a matter of amplitude amplification 
to find the temporary qubit in state $\left| 0 \right\rangle$ when a partial measurement is applied 
on that particular temporary qubit. This will erase the unnecessary states directly from the system, and will leave 
a superposition with the rest of the states.

\subsection {Performing Conditional Operations on Some Selected Records}

A usual scenario in the processing of a database is to select certain sets of records, 
each set is selected based on some condition, then apply an operation 
on the {\it intersection} of the selected set of records according some global condition. 
For example, assume that $R_1$ and $R_2$ are 
two selected set of records according to the two conditions $c_1$ and $c_2$ respectively. 
Assume that an operation $U$ should be applied on the intersection of the selected records according to 
the global condition $(c_1 \, AND\,(NOT\, c_2))$. Fig.~\ref{Selectfig} 
shows such construction where the set $R_1$ of records is selected by a Boolean function $f_1$ and 
the set $R_2$ of records is selected by a Boolean function $f_2$. Both selected records 
are combined using the global condition $(c_1 \, AND\,(NOT\, c_2))$ on the last temporary 
qubit and a conditional application of $U$ is done for only the records that {\it satisfy} 
the global condition. In general, to apply such an arbitrary operator $U$ 
on $k$ selected set of records, $k+1$ temporary qubits are required.

\begin{center}
\begin{figure}  [htbp]
\begin{center}

\setlength{\unitlength}{3947sp}%
\begingroup\makeatletter\ifx\SetFigFont\undefined%
\gdef\SetFigFont#1#2#3#4#5{%
  \reset@font\fontsize{#1}{#2pt}%
  \fontfamily{#3}\fontseries{#4}\fontshape{#5}%
  \selectfont}%
\fi\endgroup%
\begin{picture}(4810,2300)(1580,-2784)
{\thinlines
\put(2921,-2229){\circle{134}}
}%
{\put(5563,-2709){\circle*{134}}
}%
{\put(4840,-2701){\circle{134}}
}%
{\put(4838,-2463){\circle*{134}}
}%
{\put(4839,-2227){\circle*{134}}
}%
{\put(5078,-2700){\circle{134}}
}%
{\put(5078,-2228){\circle*{134}}
}%
{\put(4231,-2468){\circle{134}}
}%
{\put(3721,-1842){\framebox(1013,1325){}}
}%
{\put(2119,-580){\line( 1, 0){296}}
}%
{\put(2126,-813){\line( 1, 0){296}}
}%
{\put(2426,-1836){\framebox(1013,1325){}}
}%
{\put(2126,-1759){\line( 1, 0){296}}
}%
{\put(4741,-579){\line( 1, 0){296}}
}%
{\put(4744,-815){\line( 1, 0){296}}
}%
{\put(4736,-1759){\line( 1, 0){296}}
}%
{\put(3440,-576){\line( 1, 0){271}}
}%
{\put(3448,-811){\line( 1, 0){271}}
}%
{\put(3448,-1763){\line( 1, 0){271}}
}%
{\put(2125,-2229){\line( 1, 0){2932}}
}%
{\put(2169,-496){\line(-1, 0){116}}
}%
{\put(2053,-496){\line( 0,-1){1331}}
\put(2053,-1827){\line( 1, 0){120}}
}%
{\put(2137,-2462){\line( 1, 0){2932}}
}%
{\put(5039,-1826){\framebox(1013,1325){}}
}%
{\put(2143,-2701){\line( 1, 0){2932}}
}%
{\put(4993,-2228){\line( 1, 0){1368}}
\put(6361,-2228){\line( 0, 1){  0}}
}%
{\put(4999,-2462){\line( 1, 0){1368}}
\put(6367,-2462){\line( 0, 1){  0}}
}%
{\put(4994,-2701){\line( 1, 0){1368}}
\put(6362,-2701){\line( 0, 1){  0}}
}%
{\put(6060,-578){\line( 1, 0){312}}
\put(6372,-578){\line( 0, 1){  0}}
}%
{\put(6060,-823){\line( 1, 0){312}}
\put(6372,-823){\line( 0, 1){  0}}
}%
{\put(6066,-1762){\line( 1, 0){312}}
\put(6378,-1762){\line( 0, 1){  0}}
}%
{\put(5564,-2771){\line( 0, 1){936}}
}%
{\put(4835,-2164){\line( 0,-1){600}}
}%
{\put(5077,-2168){\line( 0,-1){600}}
}%
{\put(2917,-2296){\line( 0, 1){458}}
}%
{\put(4231,-2532){\line( 0, 1){696}}
}%
\put(1558,-966){\makebox(0,0)[lb]{\smash{{\SetFigFont{12}{14.4}{\rmdefault}{\mddefault}{\updefault}{$n$}%
}}}}
\put(1395,-1125){\makebox(0,0)[lb]{\smash{{\SetFigFont{12}{14.4}{\rmdefault}{\mddefault}{\updefault}{qubits}%
}}}}
\put(6268,-1291){\makebox(0,0)[lb]{\smash{{\SetFigFont{12}{14.4}{\rmdefault}{\mddefault}{\updefault}{$\vdots$}%
}}}}
\put(2137,-1284){\makebox(0,0)[lb]{\smash{{\SetFigFont{12}{14.4}{\rmdefault}{\mddefault}{\updefault}{$\vdots$}%
}}}}
\put(4112,-1215){\makebox(0,0)[lb]{\smash{{\SetFigFont{12}{14.4}{\rmdefault}{\mddefault}{\updefault}{$U_{f_{2}}$}%
}}}}
\put(2819,-1202){\makebox(0,0)[lb]{\smash{{\SetFigFont{12}{14.4}{\rmdefault}{\mddefault}{\updefault}{$U_{f_{1}}$}%
}}}}
\put(5488,-1215){\makebox(0,0)[lb]{\smash{{\SetFigFont{12}{14.4}{\rmdefault}{\mddefault}{\updefault}{$U$}%
}}}}
\put(450,-2716){\makebox(0,0)[lb]{\smash{{\SetFigFont{12}{14.4}{\rmdefault}{\mddefault}{\updefault}{$(c_1 \, AND\,(NOT\, c_2))$}%
}}}}
\put(1913,-2482){\makebox(0,0)[lb]{\smash{{\SetFigFont{12}{14.4}{\rmdefault}{\mddefault}{\updefault}{$c_2$}%
}}}}
\put(1917,-2242){\makebox(0,0)[lb]{\smash{{\SetFigFont{12}{14.4}{\rmdefault}{\mddefault}{\updefault}{$c_1$}%
}}}}
\end{picture}%

\end{center}
\caption{Conditioal application of an arbitrary operation $U$ based on two SELECT operators, where $c_1 c_2  \oplus c_1  \equiv c_1 \,AND\,(NOT\,c_2 )$.}
\label{Selectfig}
\end{figure}
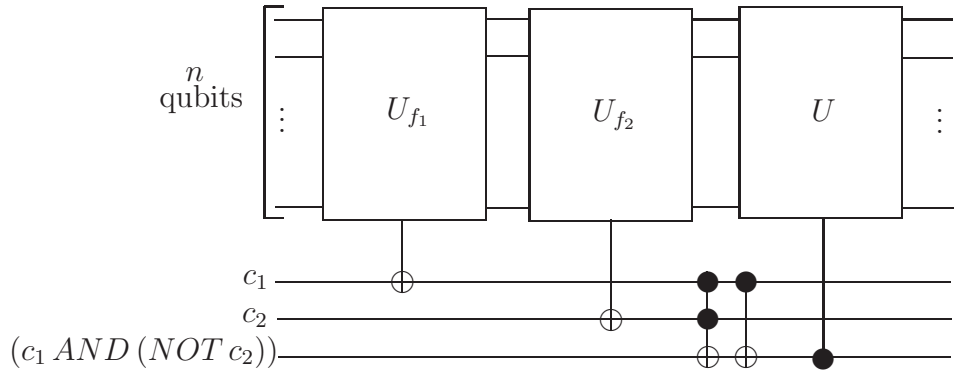
\end{center}

\subsection{Backing Up a Required Portion of a Database File (BACKUP)}

\begin{center}
\begin{figure}  [htbp]
\begin{center}
\setlength{\unitlength}{3947sp}%
\begingroup\makeatletter\ifx\SetFigFont\undefined%
\gdef\SetFigFont#1#2#3#4#5{%
  \reset@font\fontsize{#1}{#2pt}%
  \fontfamily{#3}\fontseries{#4}\fontshape{#5}%
  \selectfont}%
\fi\endgroup%
\begin{picture}(3472,1971)(1580,-2455)
{\thinlines
\put(2921,-2229){\circle{134}}
}%
{\put(2119,-580){\line( 1, 0){296}}
}%
{\put(2126,-813){\line( 1, 0){296}}
}%
{\put(2426,-1836){\framebox(1013,1325){}}
}%
{\put(2126,-1759){\line( 1, 0){296}}
}%
{\put(4741,-579){\line( 1, 0){296}}
}%
{\put(4744,-815){\line( 1, 0){296}}
}%
{\put(4736,-1759){\line( 1, 0){296}}
}%
{\put(3440,-576){\line( 1, 0){271}}}

{\put(3448,-811){\line( 1, 0){271}}}

{\put(3448,-1763){\line( 1, 0){271}}}

\put(2176,-1386){$\vdots$}

\put(4876,-1386){$\vdots$}

{\put(2918,-2289){\line( 0, 1){458}}
}%
{\put(2169,-496){\line(-1, 0){116}}
}%
{\put(2053,-496){\line( 0,-1){1331}}
}
{\put(2053,-1827){\line( 1, 0){120}}
}%
{\put(3715,-2368){\framebox(1026,1860){}}
}%
{\put(2124,-2227){\line( 1, 0){1584}}
}%
{\put(4748,-2227){\line( 1, 0){288}}
}%
\put(2854,-1217){\makebox(0,0)[lb]{\smash{{\SetFigFont{12}{14.4}{\rmdefault}{\mddefault}{\updefault}{$U_f$}%
}}}}
\put(858,-966){\makebox(0,0)[lb]{\smash{{\SetFigFont{12}{14.4}{\rmdefault}{\mddefault}{\updefault}{superposition }%
}}}}
\put(858,-1125){\makebox(0,0)[lb]{\smash{{\SetFigFont{12}{14.4}{\rmdefault}{\mddefault}{\updefault}{of $n$ qubits}%
}}}}
\put(1638,-2277){\makebox(0,0)[lb]{\smash{{\SetFigFont{12}{14.4}{\rmdefault}{\mddefault}{\updefault}{$\left| 0 \right\rangle$}%
}}}}
\put(1354,-2081){\makebox(0,0)[lb]{\smash{{\SetFigFont{12}{14.4}{\rmdefault}{\mddefault}{\updefault}{extra qubit}%
}}}}
\put(4132,-1194){\makebox(0,0)[lb]{\smash{{\SetFigFont{12}{14.4}{\rmdefault}{\mddefault}{\updefault}{$D_p$}%
}}}}
\end{picture}%
\end{center}
\caption{Backing up a portion of a database file.}
\label{backupfig}
\end{figure}
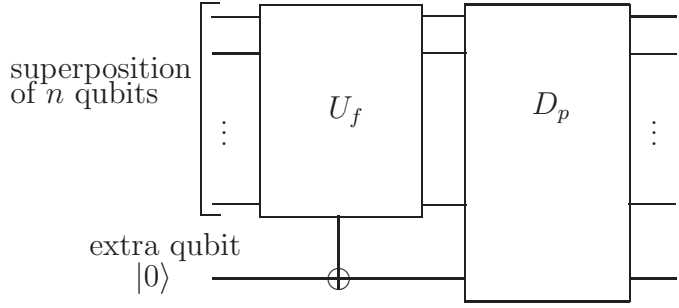
\end{center}

Suppose that a copy of some states in a superposition should be stored in a safe to be protected 
from any arbitrary operations to be done by mistake on the superposition. 
To achieve this, assume that $f$ is a Boolean function that 
identifies the records to be backed up. Firstly, apply $U_f$ on the superposition 
taking a temporary qubit as the target qubit, this creates an entanglement between 
the required subspace and the temporary qubit in state 
$\left| 1 \right\rangle$, and the rest of the system entangled 
with the temporary qubit in state $\left| 0 \right\rangle$. This temporary qubit 
will be considered as the {\it key} of the safe (the safe key).   

Now, there are two separate subspaces in the superposition. A subspace entangled with the temporary qubit 
in state $\left| 1 \right\rangle$ representing the items sent to the backup and the 
rest of the superposition that doesn't contain the states in the backup, 
entangled with state $\left| 0 \right\rangle$ of the temporary qubit. To create a copy of 
the states in the backup and insert them in the subspace entangled with state $\left| 0 \right\rangle$ 
, apply the partial diffusion operator $D_p$ on the system including 
the temporary qubit. The mechanism of these operations can be understood as follows: 
Assume that the system is initially as follows,

\begin{equation}
\label{ENheq11}
\left| \psi_0  \right\rangle
\sum\limits_{i = 0}^{2^n  - 1} {\alpha _i \left| i \right\rangle  \otimes \left| 0 \right\rangle}.
\end{equation}

\begin{itemize}
\item[1-] {\it Applying the Oracle}. Apply the oracle $U_{f}$ that maps the items in the list to either 
0 or 1 simultaneously and stores the result in the temporary qubit:

\begin{equation}
\label{ENheq12}
\begin{array}{l}
\left| \psi_1  \right\rangle= U_f \left| \psi_0  \right\rangle  = U_f \sum\limits_{i = 0}^{2^n  - 1} {\alpha _i \left| i \right\rangle  \otimes \left| 0 \right\rangle  = } \sum\limits_{i = 0}^{2^n  - 1} {\alpha _i \left| i \right\rangle  \otimes \left| {f(i)} \right\rangle }.  \\ 
 \end{array}
\end{equation}


\item[2-]{\it Partial Diffusion}. Let $M$ be the number of matches, which make the oracle $U_f$ 
evaluate to true, i.e. items to be sent to the backup and $N=2^n$. 
Assume that $\sum\nolimits_i {{'}} $ denotes a sum over $i$ representing the items to be sent to the backup, 
and $\sum\nolimits_i {{''}} $ denotes a sum over $i$ representing the rest of the items in the list. 
So, the system  $\left| \psi_1 \right\rangle$ shown in Eqn.~(\ref{ENheq12}) can be written as follows:

\begin{equation}
\label{ENheq16}
\left| \psi_1  \right\rangle = \sum\limits_{i = 0}^{N - 1} {^{''} \alpha _i \left( {\left| i \right\rangle  \otimes \left| 0 \right\rangle } \right)}  + \sum\limits_{i = 0}^{N - 1} {^{'} \alpha _i \left( {\left| i \right\rangle  \otimes \left| 1 \right\rangle } \right)}. 
\end{equation}

Applying $D_p$ on $\left| \psi_1  \right\rangle$ will result in a new system described as follows: 

\begin{equation}
\label{ENheq17}
\left| \psi_2  \right\rangle  = \sum\limits_{i = 0}^{N - 1} {^{''}a_i \left( {\left| i \right\rangle  \otimes \left| 0 \right\rangle } \right)}  +  \sum\limits_{i = 0}^{N - 1} {^{'} b_i\left( {\left| i \right\rangle  \otimes \left| 0 \right\rangle } \right)}  +  \sum\limits_{i = 0}^{N - 1} {^{'} c_i \left( {\left| i \right\rangle  \otimes \left| 1 \right\rangle } \right)}, 
\end{equation}

\noindent
where the mean used in the definition of partial diffusion operator is,

\begin{equation}
\label{ENheq18}
\left\langle \alpha  \right\rangle  = \frac{1}{N}\left( {\sum\limits_{i = 0}^{N - 1} {^{''} \alpha _i } } \right),
\end{equation}

and $a_i$, $b_i$ and $c_i$ used in Eqn. \ref{ENheq17} are calculated as follows:

\begin{equation}
\label{ENheq19}
a_i  = 2\left\langle {\alpha} \right\rangle  - \alpha_i,\,\,\,\,\,\
b_i  = 2\left\langle {\alpha} \right\rangle, \,\,\,\,\,\ 
c_i  = -\alpha_i.
\end{equation}

Notice that, the states with amplitude $b_i$ had amplitude {\it zero} before applying $D_p$. The system 
ends up with a copy of the required states, previously sent to the backup by the oracle, 
in the subspace entangled with state $\left| 0 \right\rangle$ of the safe key qubit. 
Applying any further operations on the records of the database {\it should} be applied by controlling that 
operations by the temporary qubit to be in state $\left| 0 \right\rangle$, in an equivalent manner to that shown in Eqn.(\ref{ContU}), 
keeping the backup in the safe entangled with state $\left| 1 \right\rangle$ of the temporary qubit. 
Notice that, a superposition of the database file together with its backup cost 
an extra qubit added to the system. 
\end{itemize}

\subsection {Restoring a Backup}

Suppose that some required records are lost from the superposition due to some invalid 
update and/or mistaken deletion providing that, a copy of these states 
has been kept in a backup and all applied operations were controlled with 
the safe key qubit to be in state $\left| 0 \right\rangle$. So, the system can be represented as follows,

\begin{equation}
\left| {\psi^{'} } \right\rangle  = \sum\limits_{i = 0}^{N - 1} {^{''} a_{i}^{'} \left( {\left| i \right\rangle  \otimes \left| 0 \right\rangle } \right)}  + \sum\limits_{i = 0}^{N - 1} {^{'''} b_{i}^{'} \left( {\left| i \right\rangle  \otimes \left| 0 \right\rangle } \right)}  + \sum\limits_{i = 0}^{N - 1} {^{'} c_i \left( {\left| i \right\rangle  \otimes \left| 1 \right\rangle } \right)},
\end{equation}

\noindent
where $\sum\nolimits_i {{'}} $ denotes a sum over $i$ representing the items in the safe, 
and $\sum\nolimits_i {{''}} $ denotes a sum over $i$ representing the rest of the items in the list, and 
$\sum\nolimits_i {{'''}} $ denotes a sum over $i$ representing the set of the correct items left in the superposition 
after applying the invalid operations. Applying the oracle $U_f$, 
originally used to create the backup, on $\left| {\psi^{'} } \right\rangle$ 
will {\it flip} the safe key qubit only for the items in $\sum\nolimits_i {{'}}$ and $\sum\nolimits_i {{'''}} $, 
sending the remaining correct items left in the superposition to the backup safe and restoring the items 
in the safe to the superposition entangled with state $\left| 0 \right\rangle$ 
as follows,

\begin{equation}
U_f \left| {\psi ^{'} } \right\rangle  = \sum\limits_{i = 0}^{N - 1} {^{''} a_{i}^{'} \left( {\left| i \right\rangle  \otimes \left| 0 \right\rangle } \right)}  + \sum\limits_{i = 0}^{N - 1} {^{'} c_i \left( {\left| i \right\rangle  \otimes \left| 0 \right\rangle } \right)}  + \sum\limits_{i = 0}^{N - 1} {^{'''} b_{i}^{'} \left( {\left| i \right\rangle  \otimes \left| 1 \right\rangle } \right)}. 
\end{equation}

Since the items 
in the backup safe is no longer valid (as a set of items), they can be deleted by 
the DELETE operator. A new fresh backup could be created using the BACKUP operator.

\section{Conclusion}

The quantum databases are expected to replace the classical databases once quantum computers are implemented on the commercial scale. Quantum computers can behave classically if a superposition is not used. Superposed quantum database will be useful in reducing the processing time where many operations could be done simultaneously on a database file as well as saving memory space. Extracting useful information from a quantum computer in a superposition is still under investigation by many researchers. Distributed processing of databases could be possible where teleportation might help in sending a quantum database file in a superposition from one place to another instantly for further processing and extracting useful information.

The QQL operators defined in this paper still require further investigation to adjust the amplitudes of the system as required. General purpose amplitude manipulation techniques must be found to be combined with the operators of the QQL. Finding a quantum version of referential integrity and relational algebra to get useful information from larger databases where many database files are used could be the next research step.

To summarize, in this paper, a method for inserting exponential number of items simultaneously as well as inserting item-by-item to a superposition has been defined. A method to update many records simultaneously has been shown. A way to delete certain records from the database simultaneously has been suggested which still need special attention as a separate problem. Performing the selection of some records and applying conditional operations on the intersection of these selected records has been shown. And finally a method to backup and restore a database file without the need of vast extra memory has been proposed.

\bibliography{QQL}
\bibliographystyle{plain}

\end{document}